Characteristic Characteristics

Authors: Sean Brocklebank[a], Scott Pauls [b], Daniel Rockmore [b,c] and Timothy C. Bates[d,e]

[a] School of Economics, University of Edinburgh

[b] Department of Mathematics, Dartmouth College

[c] The Santa Fe Institute

[d] Department of Psychology, University of Edinburgh

[e] Centre for Cognitive Aging and Cognitive Epidemiology, University of Edinburgh

Correspondence:

Sean Brocklebank

sean.brocklebank@ed.ac.uk

School of Economics

University of Edinburgh

Edinburgh EH8 9JZ

United Kingdom

Abstract

While five-factor models of personality are widespread, there is still not universal agreement on this as a structural framework. Part of the reason for the lingering debate is its dependence on factor analysis. In particular, derivation or refutation of the model via other statistical means is a worthwhile project. In this paper we use the methodology of spectral clustering to articulate the structure in the dataset of responses of 20,993 subjects on a 300-item item version of the IPIP NEO personality questionnaire, and we compare our results to those obtained from a factor analytic solution. We found support for five- and six-cluster solutions. The five-cluster solution was similar to a conventional five-factor solution, but the six-cluster and six-factor solutions differed significantly, and only the six-cluster solution was readily interpretable: it gave a model similar to the HEXACO model. We suggest that spectral clustering provides a robust alternative view of personality data.

## 1. Introduction

The five-factor model (FFM) of personality has become "the default model of personality structure" (McCrae & Costa, 2008, p.273). While this is may very well be true, the FFM is by no means the only extant model of personality; models with six (Ashton & Lee, 2007), three (Eysenck 1991, 1992a) two (Digman, 1997) and even one factor (Musek, 2007) have been proposed and studied. Within five-factor models, debates remain about the allocation of facets under domains (Backstrom, Larsson & Maddux, 2009), and clustering of facets into "aspects" under factors (DeYoung, Quilty, & Peterson, 2007). In large part debates regarding the structure of personality will require information outside statistical modelling of items (Eysenck, 1992b). However modelling per se can cast light on the relative fit of alternative models.

Factor analysis (FA) and, more recently confirmatory factor analysis (CFA: Jöreskog, 1969) remain the primary techniques for exploring structure in questionnaire data and are the foundation for the FFM. Attempts to confirm the FA derivation of the FFM have met with mixed success (Marsh et al, 2010; Gignac, Bates & Jane, 2007). Alternatives to date are restricted to one recent paper that did not employ factor analysis or its near-relations: Tiliopoulos, Pallier and Coxon (2010) used non-metric multidimensional scaling to examine the structure of a five-factor personality questionnaire. Here we use spectral clustering (Von Luxburg, 2007) to examine data from Johnson (2005), comprising responses from 20,993 subjects to a 300-item questionnaire assessing the five-factor model. These results are compared to those of a corresponding factor analysis. Our study is, to the best of our knowledge, the first application of spectral clustering to a personality dataset. We first outline spectral clustering, then introduce its application to personality.

## 1.1 Spectral Clustering

Spectral clustering is one of a family of clustering techniques that share with FA the basic objective of creating a low-dimensional representation of the data. Where the techniques differ is in the specific problem they are optimizing. Factor analysis maximizes the amount of variance that can be accounted for in a given low-dimensional projection of the correlation matrix (Spearman 1924, Cattell 1978). Spectral clustering, on the other hand, minimizes the amount of cutting necessary to divide a geometric representation of the data into separate clusters. More formally, spectral clustering is a network-based analysis technique which, by using a network representation of the data (summarized as weighted connections between nodes), finds a representation of the network as a union of disjoint subsets that effectively minimizes the overall interpoint weights between clusters, while simultaneously accounting for the relative sizes of the clusters. In our case, the *nodes* of the network are indexed by the 300 questions and the connections (*edges*) between them are weighted as a function of the correlation between the corresponding questions — strong positive correlations translate to large weight and strong negative correlations translate to small weight, and weak correlations correspond to an intermediate weight. Details for this procedure are in section 3.

Generally speaking, the goal of clustering for a network ("community finding" as it is sometimes called) is to find, for a given value of $k$, the decomposition of the network as a disjoint union of $k$ subnetworks, such that the overall weight between the clusters is minimized. The minimization usually accounts for or normalizes by a factor that accounts for either the size of the sub-networks (RatioCut) or the density (NCut) where density is computed as the sum of the "degrees" (the degree of a node is the sum of the weights of the edges incident at that node) of the nodes in the cluster. The exact solution of this optimization problem is prohibitive (exponential in the square of the number of nodes), but a "relaxation" of the discrete problem to a continuous problem with similar constraints produces an exactly solvable eigenvector problem. Translation back to the discrete setting arrives at a very good approximation to an exact solution to the original combinatorial problem (Shi & Malik, 2002). Each cluster then represents a coherent subset of the network, with relatively high density within and relatively low connectivity without, and thus in our case, a subset of nodes with a relatively high degree of commonality and explanatory power. Each cluster can effectively be viewed as a summary dimension for the underlying data (see e.g., Leibon et al., 2008).

While both spectral clustering and traditional factor analysis have the same broad goal – to uncover a small number of dimensions (factors) which explain the data, the methodologies are quite different. Traditional factor analysis attempts to find factors and loadings which maximize the explained variance in the data. Spectral clustering attempts to find geometrically optimal clusters and while these clusters may be used to form the analogue of factors (and consequently loadings), the optimization criteria do not include conditions minimizing residual variance, but rather a minimization of the weights of the severed connections.

2. Method

2.1 Participants and procedure

A total of 23,994 people completed the online 300-item IPIP representation of the NEO PI-R between August 6, 1999 and March 18, 2000. Only 20,993 of these submissions were used after excluding subjects for long strings of identical or missing responses, or for duplicate submissions. Detailed information about the criteria for excluding responses can be found in Johnson (2005). The final sample was 63.1% female and had a mean age of 26.1 years (SD=10.7 years). Subjects were not actively recruited; they discovered the website either on their own or through word-of-mouth.

2.2 Measures

Subjects completed a 300-item International Personality Item Pool (IPIP) representation of the NEO PI-R. As noted in Goldberg et al. (2006), the IPIP proxies have been developed to measure the same constructs as their proprietary equivalents. For the NEO, this includes five domain-level constructs: Neuroticism (N),

Extraversion (E), Openness (O), Agreeableness (A), and Conscientiousness (C). The IPIP NEO also measures six facets per domain. For this implementation the mean correlation of facets on the IPIP proxy with corresponding facets in the NEO PI-R is 0.94 after correcting for unreliability (Goldberg, 1999).

## 3. Results

Approximately 0.5% of the data were missing. For analytical purposes, missing entries were treated as "neither agree nor disagree."

### 3.1 Spectral clustering

#### 3.1.1 Construction of the network

To create the undirected network, we construct a symmetric adjacency matrix, $A$, where $a_{ij} = a_{ji}$ is the weight of the edge between node $i$ and node $j$ (recall that each node represents one of the original 300 questions) derived from the correlation matrix on the items as follows: First the correlations are transformed to Euclidean distances on a unit sphere:

$d_{ij} = [(1-c_{ij})/2]^2$

The resulting matrix is a dissimilarity matrix as the distances are larger if the items are dissimilar (with respect to correlation). We obtain a similarity matrix, to use as our weighted adjacency matrix, by applying a Gaussian to our dissimilarity measures:

$a_{ij} = \exp(-d_{ij}/\sigma)^2$

The parameter $\sigma$ is a scale parameter – it determines how wide the Gaussian is. The effect of this is to specify how small the distances must be to create a strong edge in the resulting adjacency matrix. Recall that an edge strength of 1 is the strongest edge while 0 denotes the absence of an edge. To see roughly how this works, consider different values of $d_{ij}$ compared to a fixed $\sigma$. If $d_{ij}$ is much smaller than $\sigma$, then $a_{ij}$ is close to 1. Conversely if $d_{ij}$ is much larger than $\sigma$, then aij is close to 0. With our data, Figure 1 shows, on the left, the histograms of entries of $a_{ij}$ with three different choices of $\sigma$. On the right, we see images of the A matrices themselves – darker colors are close to zero while lighter colors are closer to 1. The effect is, hopefully, clear – smaller $\sigma$ create smaller $a_{ij}$ while larger $\sigma$ create larger $a_{ij}$.

---------- Insert Figure 0 about here ----------

#### 3.1.2 Spectral clustering algorithm

Spectral clustering relies on an analysis of the spectral data (eigenvalues and eigenvectors) of the Laplacian. We form the *symmetrized graph Laplacian* from the adjacency matrix $A$,

$$L = I - D^{-1/2}AD^{-1/2}.$$

Here, $I$ is the identity matrix and $D$ is the diagonal matrix of *degrees* of the nodes defined by,

$$d_i = \sum_{j=1}^{300} a_{ij}.$$

The "NCut problem" for the weighted network encoded in $A$ is to find a decomposition of the nodes of the network into disjoint subsets $A_1,\dots,A_k$ (for given $k$) that minimizes the sum

Cut($A_1$)/Vol($A_1$) + … +Cut($A_k$)/Vol($A_k$)

where Cut($A_i$) is the total weight of edges from nodes in $A_i$ out to the rest of the network and Vol($A_i$) is the sum of the degrees in the nodes of $A_i$. As shown in (Shih and Malik, 2000) a relaxation of this combinatorial problem turns this into an eigenvector problem for this symmetrized Laplacian. To accomplish this we proceed as follows: (see Ng, Jordan and Weiss, 2002):

1. Find the eigenvalues and eigenvectors of the symmetrized Laplacian. It is possible to show that the matrix is positive semidefinite – i.e., has only nonnegative eigenvalues. Discard the zero eigenvalues and associated eigenvectors.

2. Determine the number of eigenvectors, $l$, we wish to use for the analysis.

3. Determine the number of clusters, $k$, we wish to use for the analysis.

4. Use $k$-means clustering on the embedding of the nodes into $\mathbb{R}^l$ given by the first $l$ (undiscarded) eigenvectors.

### 3.1.3 Parameter estimation

As detailed above, the spectral clustering algorithm uses three parameters: the number of eigenvectors to feed into the clustering algorithm ($l$), the scale parameter ($\sigma$), and the number of clusters ($k$). The success of the method depends on picking reasonable choices for these parameters. While we approach picking these parameters sequentially, it is important to realize that all three are intertwined. For example, at different scales, we may find different good choices for $l$ and $k$. In a sense, we wish to pick values for these three parameters in conjunction with one another.

### 3.1.3.1 The number of eigenvectors

In contrast to other spectral methods (such as principal components analysis), the smaller the eigenvalue of the symmetrized Laplacian, the more important it is. The zero eigenvalues correspond to the connected components (two nodes are connected if there is a sequence of nonzero weight links between the nodes and a connected component is a maximally connected subnetwork) of the networks, naturally thought of as the most obvious form of clustering. As the eigenvalues increase (i.e., move away from zero), they encode a diminishing amount of clustering information. If the spectrum naturally separates into a clump of eigenvalues near 0 and a clump separate from that, the "lower" eigenvalues and corresponding eigenvectors give a natural dimensionality reduction in the data.

In picking the number of significant eigenvalue/eigenvector pairs, we use an ad hoc method similar in many respects to the "scree" method used to pick the number of factors in factor analysis. We first note that this will depend on the parameter σ which we vary between 0 and 1. The goal is to look for domains of stability in the spectrum of the symmetrised Laplacian, as measured by the number of small nonzero eigenvalues that are well separated from the "bulk" of the spectrum which generally tend to cluster around 1. Figure 1 shows a plot of the nonzero eigenvalues for a range of scale parameters σ from 0.35 to 1. We see that for a significant range of σ, the first four eigenvalues are visually separated from the bulk. We note that for σ less than 0.35, the graph associated to $A$ has more than one connected component. Thus, we will pick $l$=4 and constrain σ to fall in the range we have considered.

---------- Insert Figure 1 about here ----------

### 3.1.3.2 The scale parameter

To further estimate σ, we consider the effect of the choice of σ on the optimal number of clusters. To assess this for a given σ, we use the method of "cluster consistency" described as follows. For a fixed $l$ and σ, apply $k$-means repeatedly with $k$=2,3,…,$k_{max}$ (where $k_{max}$ is a reasonable choice for the maximum number of clusters -- for this data, we used $k_{max}$=10) to the $l$-dimensional spectral coordinates. For a given $k_p$, we then use the following procedure to test the consistency of the clustering with $k_p$ clusters we have computed above.

1. Pick 150 questions at random from the total and perform spectral clustering with $l$ fixed and $k$= $k_p$.

2. Compute the percentage of the 150 questions whose cluster classification differs from the original clustering.

In general we repeat this many times and compare the distributions of reclassifications for all $k$ to select the most consistent clustering (i.e. the $k$ which has the lowest proportion of reclassified items). We denote by $n$ the number of times we repeat the procedure.

In our case, for each σ ∈ {0.1, 0.15, 0.2,…,1} we use this procedure with $l$=4 and computed consistency for $k$=2,…,10 with 100 repetitions of the procedure above for each $k$. Figure 2 shows the mean of the minimum classification error percentages for all the trials for a given *(σ,k)* pair. In the figure, a cell has a star if it achieves the minimum misclassification error in that row (two stars are used if the cells are not significantly different from one another). Figure 3 is closely related—it plots the minimum misclassification error as a function of *σ* (i.e. it connects the stars). Figure 3 shows that classifications become more stable as *σ* increases up to a value of about 0.4, at which point it levels off with about 30% of items being misclassified. This suggests that we should choose a value of $\sigma$ in the range 0.4 to 1. Referring back to Figure 2, we see that this range appears to be most consistent a number of clusters—$k$—equal to five or six. The next section will consider the optimal value of $k$ in more depth.

---------- Insert Figure 2 about here ----------

---------- Insert Figure 3 about here ----------

### 3.1.3.3 The number of clusters

Last, we wish to fix $k$. This is in some sense the key point of the paper as the number of clusters is the analogue for the number of factors. As noted above, the optimal $k$ depends on the value of the scale parameter $\sigma$. For reasonable values of $\sigma$ (above 0.4), $k=5$ or $6$ has the lowest classification error for all values of $k$ between 2 and 10, and this holds across a range of values of the scale parameter $\sigma$. To check that these are global and not merely local optima, we repeat the method of cluster consistency for larger values of $k$ (we test $k$ up to 40) and with more repetitions per $k$ (200, instead of the previous 100). We set $l=4$, $k_{max}=40$, $n=200$, and $\sigma$ to each of $0.4$, $0.5$ and $0.75$. We set $k_{max}$ to ensure that the largest value of $k$ included in our tests substantially exceeded the number of postulated facets (30). Figure 4 shows the results. In each pane, the the horizontal axis shows the values of $k$ while the vertical shows the percentage of misclassifications. Each circle represents one trial. The solid red line is the mean while the dotted lines are the mean plus or minus one standard deviation. The top, middle and bottom panes show the results for $\sigma=0.4$, $0.5$ and $0.75$ respectively.

Figure 4 confirms that our optima are indeed global, and thus that the number of clusters is five or six. Which of these values is favoured depends, as noted above, on the choice of $\sigma$. But it is worth stressing that both values of $k$ are "correct" and both are reflected in the data. This issue is considered further in the discussion.

While we do not focus here on the facet-level of personality, it is noteworthy that while Figure 3 shows evidence for structure at the domain level (i.e., five major clusters), there is no indication of facet-level structure within the data. Particularly, there is no obvious drop in the proportion of reclassified items around 30 clusters, which is what we would have expected if there had been measurable structure at the facet level.

---------- Insert Figure 4 about here ----------

---------- Insert Table 1 about here ----------

3.2 Comparison of domains derived from spectral clustering and factor analysis

Since the stability analysis in section 3.1.3.3 indicates that a five- or six-cluster solution fits best, we compared the results for five- and six-cluster solutions to those obtained from varimax-rotated factor analysis. In each case, the factor solution was calculated by assigning each item to the factor on which the item has its highest loading in a varimax rotation. The cluster solution is calculated as the configuration of item assignments with the lowest sum of squares out of 10,000 runs of the spectral clustering algorithm described in section 3.1.

Table 2 compares the allocation of the 300 test items by a five-factor solution versus a five-cluster solution. The rows of the table denote which cluster them items were assigned to, while columns denote which factor items were assigned to. When items appear on the main diagonal, this means that they were assigned similarly by the two methods. Empty cells imply that there were no assigned values in the category (e.g. no items from the N-factor were assigned to the A-cluster, but one item from the N-cluster was assigned to the A-factor). Note that there are only 14 off-diagonal numbers, meaning that 286/300 items were assigned similarly by the two procedures.

Similarly, Table 3 compares the allocation of the 300 test items by a six-factor solution versus a six-cluster solution. These two analyses gave sharply contrasting solutions. The FA sixth factor contained only seven items (4 from facet N4: Self-Consciousness, and 3 from other A and N facets). By contrast, the six-cluster solution yielded a large sixth cluster containing 34 items. This sixth cluster was primarily made up of A and C items and closely resembles the Honesty-Humility factor of the six-dimensional HEXACO model. Thus while the sixth *factor* is a small nuisance factor with a majority of members from a single facet, the sixth *cluster* was large and meaningful.

3.3 The six-cluster solution and the HEXACO model

The results of the current analysis cannot be compared directly with results from the HEXACO personality inventory, because the IPIP NEO PI-R and the HEXACO-PI include different items, and the NEO does not adequately sample all four facets of Honesty-Humility (Ashton & Lee, 2005). But our sixth cluster is highly similar to the HEXACO H factor. Most obvious is the fact that the sixth cluster consists almost entirely of A and C items (with 1 item from each of the N and E clusters). This matches with Lee and Ashton's observation that H's correlations with the other factors are primarily with A and C (Lee & Ashton, 2004). Moreover, the 13 out of 15 A items re-assigned to the putative-H cluster are drawn come from the A2 (Morality) and A5 (Modesty) facets of A. Given that morality and modesty are synonymous with honesty and humility, this lends face validity to the notion that our sixth cluster is Honesty-Humility. This also accords with Ashton and Lee's (2005) suggestion that A2 and A5 are the facets most closely related to

Honesty-Humility. Looking at individual items which are re-assigned to the sixth cluster reinforces this impression: "Cheat to get ahead" (A2; reversed), "Take advantage of others" (A2; reversed), "Dislike being the center of attention" (A5), and "Seldom toot my own horn" (A5). Of the 18 Conscientiousness items that were assigned to the sixth cluster, 9 came from C3: Dutifulness, which contains items such as "Keep my promises," and "Tell the truth", further reinforcing the impression that the sixth cluster is Honesty-Humility.

As predicted from the HEXACO model, the Emotionality cluster absorbed items from O3(Emotionality) having high face validity for Emotional lability: "Experience my emotions intensely," "Seldom get emotional," "Am not easily affected by my emotions," and "Experience very few emotional highs and lows." Remaining items from O3 relate to attention to emotion, and were either retained within O ("Try to understand myself") or were assigned to A ("Feel others' emotions"). The Extraversion, and Openness clusters remained much as they were in the five-dimensional solutions, as predicted by the HEXACO model.

One notable difference between HEXACO theory and our six-cluster result is that all of the items from N2: Angry Hostility remained within the Emotionality cluster, whereas the HEXACO model predicts that they should end up with the other Agreeableness items. The very large number of subjects in the present experiment suggests that these items may best be retained under emotionality.

4. Discussion

The chief claim of the present study is that spectral clustering provides a more useful perspective, at least of this dataset, than does traditional factor analysis. We argue that the two methods give comparable results for five factors/clusters, but that the six-cluster solution yields more meaningful results than the six-factor solution. This distinction matters because the current battle over the number of dimensions in personality is in large part a battle between the five-dimensional OCEAN model and the six-dimensional HEXACO model. A method that can "see" both of these hypothesized structures provides a more useful way to resolve this dispute, and spectral clustering, but not factor analysis, can do this.

The finding that SC favors a six-dimensional solution very similar to the HEXACO model—represents, to our knowledge, the first time that HEXACO structure has been found in an FFM questionnaire using any unsupervised learning algorithm. Indeed, Lee and Ashton (2008) noted that there has only ever been one English lexical study to produce a result resembling the HEXACO structure, although the structure emerges repeatedly in cross-language studies.

A secondary finding - that spectral clustering favors a five-cluster model similar to the classic FFM was not obvious a priori. Given the fact that these two statistical techniques are based on different embeddings of the data and represent the solution to different objective functions, this is a significant result. This represents a type of robustness check on the FFM, which the FFM passes. This finding is in contrast with Tiliopoulos, Pallier and Coxon (2010 – discussed below).

This perspective does come with costs: computational costs (generating the results for figure 3 took our computer two days) and intellectual costs. The additional computational costs would of course be lower for smaller datasets, but SC is inevitably more computationally intensive than FA. The intellectual costs are not so easily reduced, but for those readers who do not believe that the benefits of spectral clustering outweigh the costs for their own research, the present study remains valuable for its findings.

The only published paper to similarly use a method unrelated to FA to analyze a five-factor dataset is that of Tiliopoulos, Pallier and Coxon (2010). These authors use non-metric multidimensional scaling (NMDS) on facet level responses from 384 subjects to the NEO PI-R. A factor analysis of their data yielded five factors (their criteria for selecting five was not given), but their NMDS solution supported only three "super structures", similar to the three factors of Eysenck's PEN model (Eysenck 1991, 1992a). There are two potential origins of the apparent differences between these present result and that of Tiliopoulos, Pallier and Coxon (2010). First the present dataset is much larger (20,993 vs. 384) allowing us to conduct analysis and item- rather than facet-level. As noted above several facets load on multiple domains, confounding analyses at this level. Second and perhaps more important reason, we reverse-code all of our Neuroticism items to reduce their distance from other domains, effectively turning the scale into Emotional Stability. While this decision is unimportant in the context of principal components or factor analysis, it can be critical in analyses that are based on distance measures. To illustrate the effect of this transformation, as well as to place our results in the context of those of Tiliopoulos, Pallier and Coxon (2010), we performed NMDS on both the original data set and the reverse-coded set. Spectral clustering of the raw data gives evidence for two clusters (N and not-N) and somewhat weaker evidence for five (See Figure 5). The reason for this is that, from the point of view of correlation, neuroticism items are negatively correlated with items from all the other domains, creating a large distance between the N items and the others. In attempting to replicate this configuration with minimal error, the multidimensional scaling algorithm must then focus more attention on the positions of the red nodes than on the others. By reversing the coding of the factors, item distances between domains are made more similar. Taken together, the small n, facet-level analysis, and reversal of N before analysis seems likely to account for the differences between the present result and that of Tiliopoulos, Pallier and Coxon (2010).

---------- Insert Figure 5 about here ----------

*Figures*

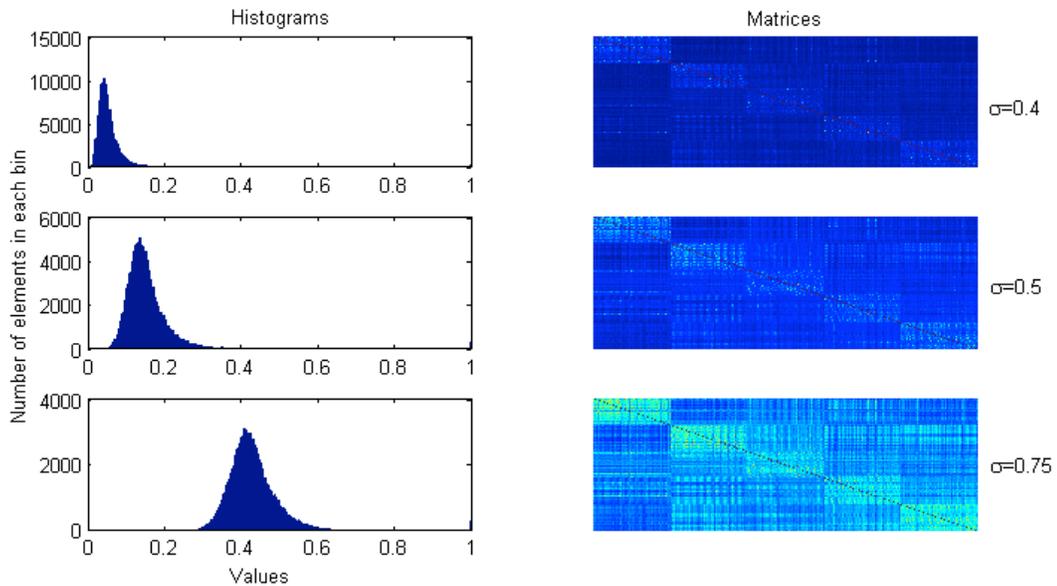

Figure 0. The effect of changing the scale parameter σ. The histograms at left show the histograms of the adjacencies $a_{ij}$ for our data as σ increases from 0.4 to 0.5 to 0.75. Notice that the adjacencies spread as the scale parameter increases. The matrices at right are images of the 300 x 300 adjacency matrices of question responses; lighter colours indicate more similar items. Note that the questions have been sorted so that all 60 neuroticism questions come before all 60 extraversion questions, and so on in the order NEOAC. Thus the block diagonal appearance of the matrices indicates the tight within-factor relationships.

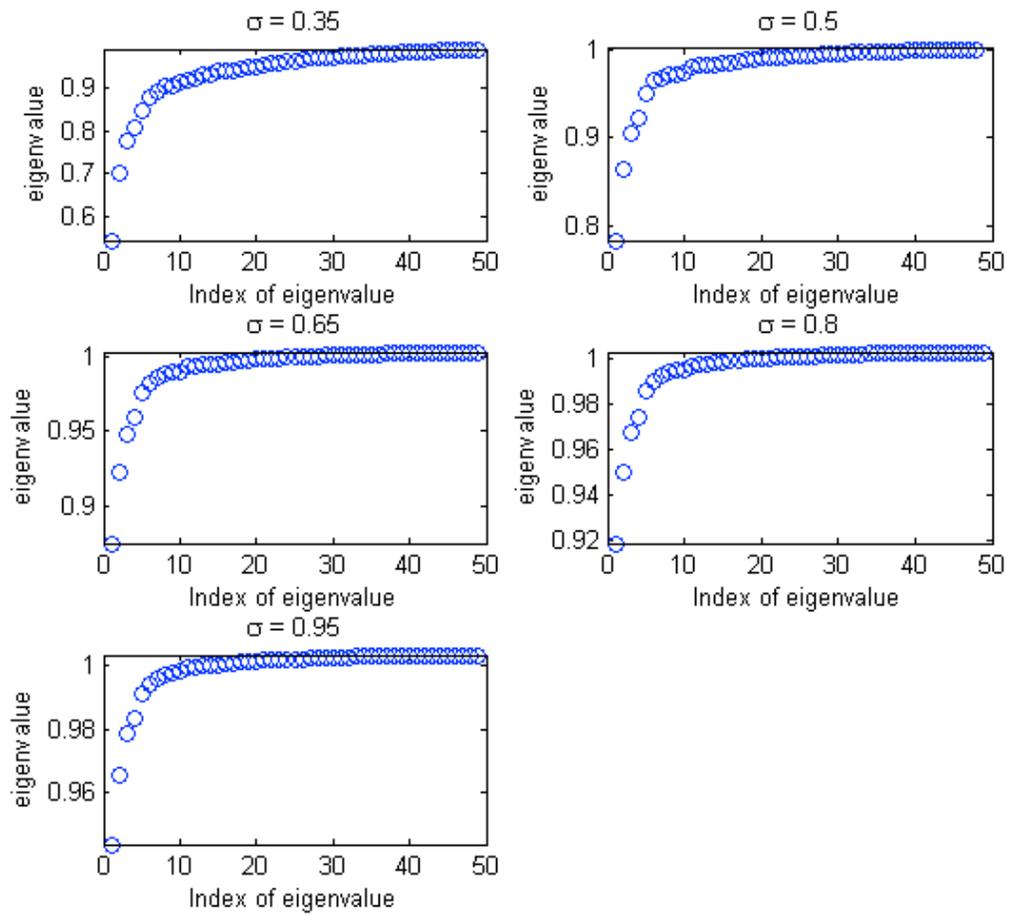

Figure 1. Eigenvalue spectra for different values of the scale parameter σ. Eigenvalues are on the vertical axis, and are indexed on the horizontal axis. The first (zero-value) eigenvalue has been omitted in all cases.

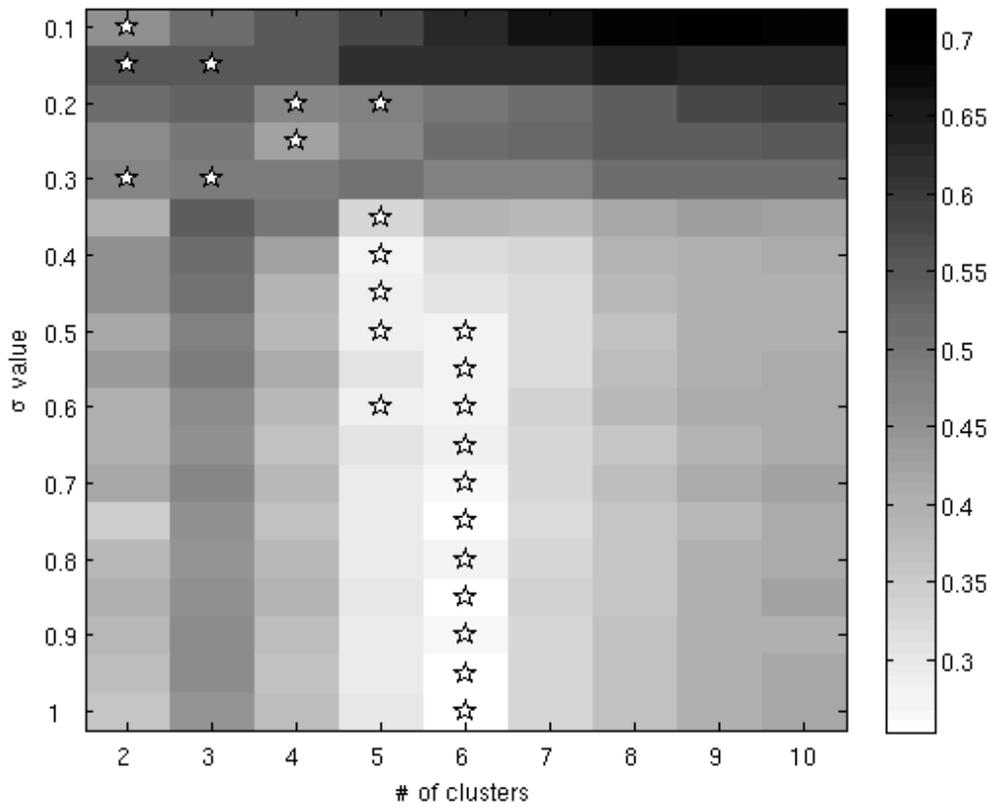

Figure 2. Mean of the minimum classification error percentages for all the trials for a given pair of k (the number of clusters) and σ (the scale parameter). Stars indicate the lowest value in a given row (multiple stars are used when the values are not significantly different from one another). Comparing the columns for 5 and 6 clusters shows that they are optimal at different levels of the scale parameter.

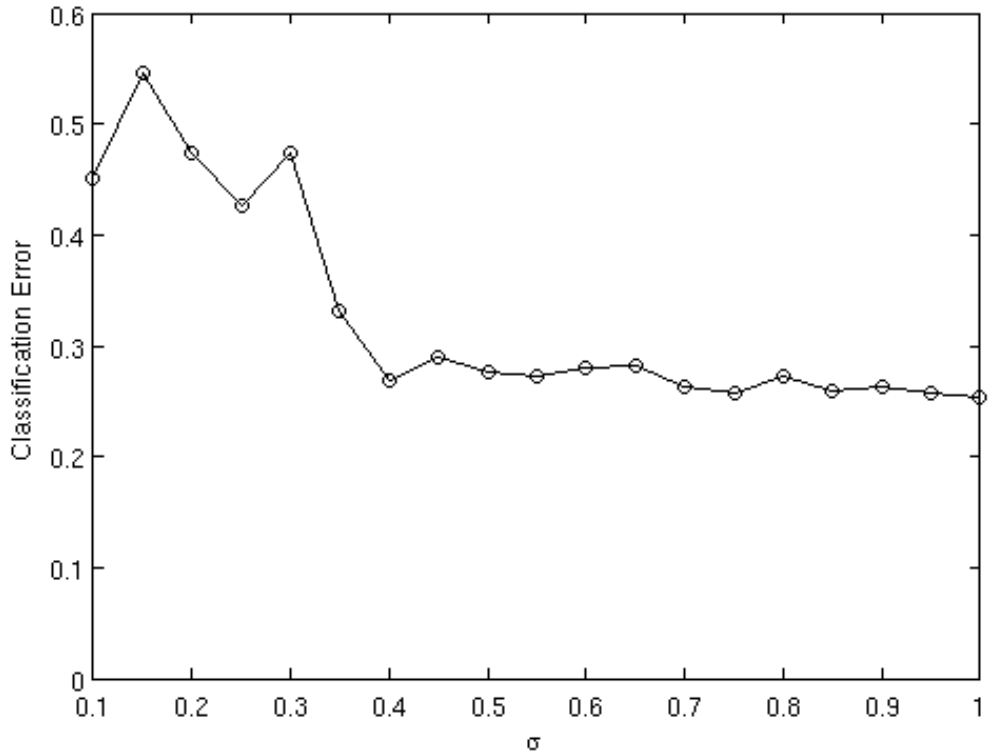

Figure 3. Minimum classification error percentages as a function of the scale parameter. This graph essentially just connects all of the stars in Figure 2. Notice that the minimum classification error is stable for values of sigma between 0.4 and 1. Comparing with Figure 2, we see that the low end of this stable range suggests five clusters, while the high end of the stable range favours six clusters. This suggests that the number of personality domains is a function of the scale at which we look at the data.

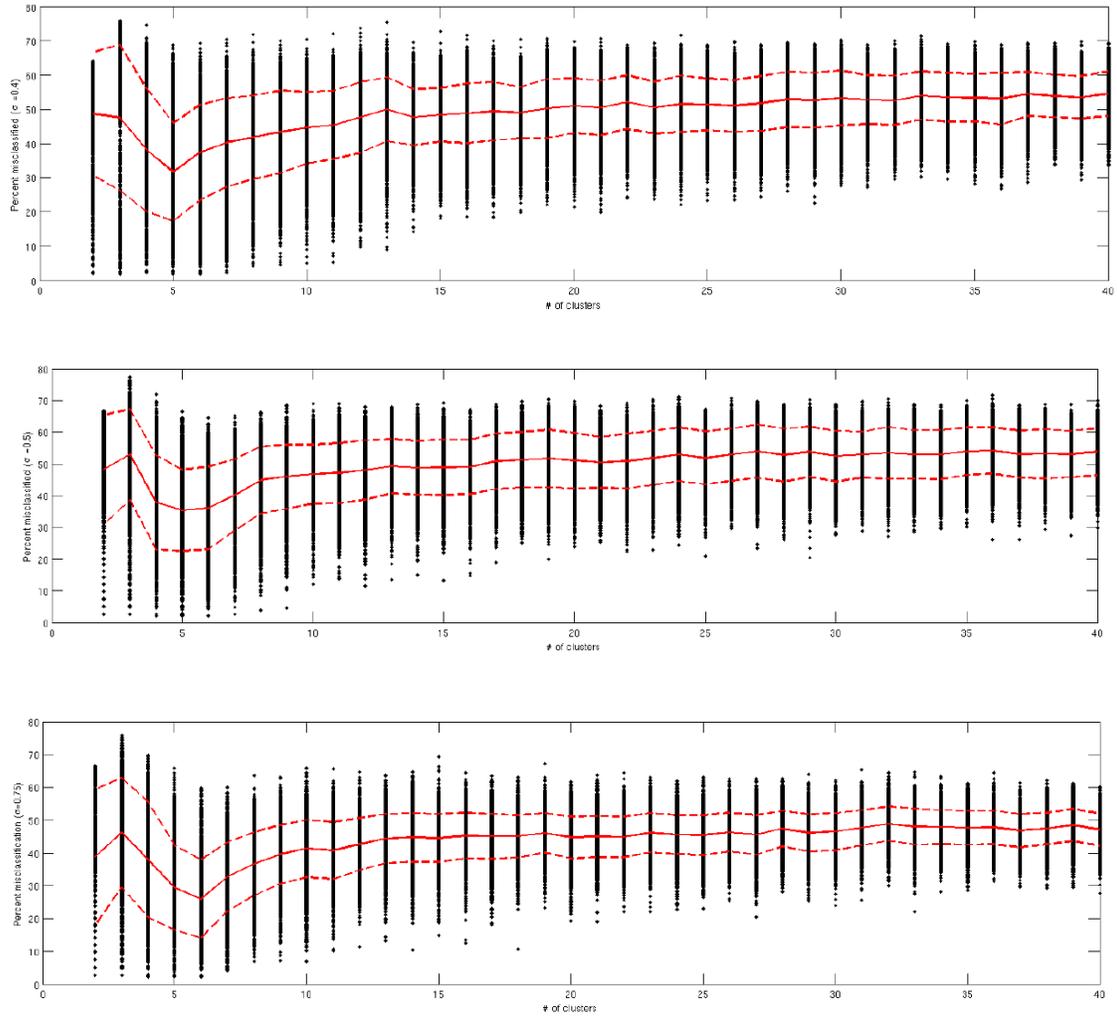

Figure 4. Proportion of questions misclassified. The vertical axis shows the proportion of items which were differentially classified by clustering on various randomly chosen 150-item subsets of the original questionnaire versus clustering based on all 300 questions. The horizontal axis shows the number of clusters. Each circle is a different trial. The solid line represents the mean misclassification while the dotted lines are plus or minus one standard deviation. The value of the scale parameter used was 0.4 in the top pane, 0.5 in the middle pane, and 0.75 at the bottom. The cluster consistency test was run 200 times for each value of $k$.

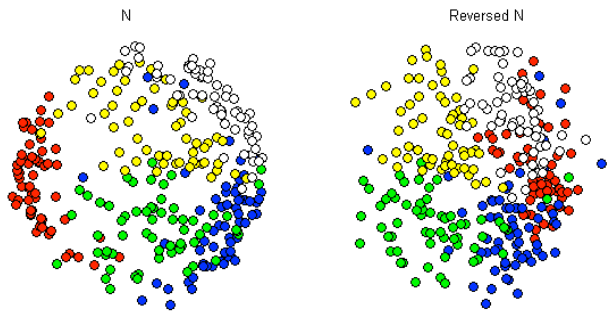

Figure 5.  Differential classification of items based on regular/reverse coding for Neuroticism. Each point is a question. Color indicates the a priori classification of the question:  green is O, white is C, blue is E, yellow is A and red is N/Reversed N. The distance between points represents their similarity: more similar items are closer.

**Tables**

Classification error when $\sigma$ = 0.4

| | Number of clusters | | | | |
|---|---|---|---|---|---|
| | k = 3 | k = 4 | k = 5 | k = 6 | k = 7 |
| mean misclassification rate | 0.48 | 0.38 | 0.32 | 0.37 | 0.4 |
| standard error | 0.0035 | 0.0028 | 0.0024 | 0.0022 | 0.0021 |

Classification error when $\sigma$ = 0.5

| | Number of clusters | | | | |
|---|---|---|---|---|---|
| | k = 3 | k = 4 | k = 5 | k = 6 | k = 7 |
| mean misclassification rate | 0.56 | 0.38 | 0.35 | 0.36 | 0.4 |
| standard error | 0.0022 | 0.0024 | 0.0021 | 0.0021 | 0.0017 |

Classification error when $\sigma$ = 0.75

| | Number of clusters | | | | |
|---|---|---|---|---|---|
| | k = 3 | k = 4 | k = 5 | k = 6 | k = 7 |
| mean misclassification rate | 0.46 | 0.38 | 0.29 | 0.26 | 0.33 |
| standard error | 0.0027 | 0.0028 | 0.0021 | 0.0019 | 0.0016 |

Table 1: Classification error as a function of sigma. This table shows the mean and standard error of the proportion of items which were misclassified for different values of σ and k. Each entry is based on 4000 simulations, and so the standard errors are small enough to ensure that all the differences within each pane are statistically significant at all conventional levels.

Table 2: Comparison of item assignment in five-cluster vs. five-factor solutions

|  | N-factor | E-factor | O-factor | A-factor | C-factor | *Cluster Size* |
|---|---|---|---|---|---|---|
| N-cluster | 65 |  |  | 1 |  | *66* |
| E-cluster |  | 46 |  | 2 |  | *48* |
| O-cluster |  | 6 | 49 |  |  | *55* |
| A-cluster |  |  |  | 63 | 4 | *67* |
| C-cluster | 1 | 5 | 1 |  | 57 | *64* |
| *Factor size* | *66* | *57* | *50* | *66* | *61* | *300* |

Table 2. This table compares the way the 300 items were classified in five-cluster vs. five-factor solutions. Rows denote which *cluster* them items were assigned to, while columns denote which *factor* items were assigned to. When items appear on the main diagonal, this means that they were assigned similarly by the two methods. Empty cells imply that there were no assigned values in the category (e.g. no items from the N-factor were assigned to the A-cluster, but one item from the N-cluster was assigned to the A-factor). Note that there are only 14 off-diagonal numbers, meaning that 286/300 items were assigned similarly by the two procedures.

Table 3: Comparison of item assignment in six-cluster vs. six-factor solutions

|            | N-factor | E-factor | O-factor | A-factor | C-factor | 6th factor | Cluster Size |
|------------|----------|----------|----------|----------|----------|------------|--------------|
| N-cluster  | 60       |          |          | 1        |          | 4          | *65*         |
| E-cluster  |          | 45       |          | 1        |          | 1          | *47*         |
| O-cluster  |          | 6        | 47       | 2        |          |            | *55*         |
| A-cluster  |          |          |          | 45       |          |            | *45*         |
| C-cluster  | 1        | 6        | 1        |          | 46       |            | *54*         |
| H-cluster  |          |          |          | 19       | 13       | 2          | *34*         |
| Factor Size| *61*     | *57*     | *48*     | *68*     | *59*     | *7*        | *300*        |

Table 3. This table compares the way the 300 items were classified in six-cluster vs. six-factor solutions. Rows denote which *cluster* them items were assigned to, while columns denote which *factor* items were assigned to. When items appear on the main diagonal, this means that they were assigned similarly by the two methods. Empty cells imply that there were no assigned values in the category (e.g. six items from the E-factor were assigned to the C-cluster, but no items from the E-cluster was assigned to the C-factor). Note that there are only 55 off-diagonal numbers, meaning that 245/300 items were assigned similarly by the two procedures.